# On the BH-galaxy relation of AGN and NLS1

**Amri Wandel[1]**

*Racah Imst. of Physics, The Hebrew University of Jerusalem*
*E-mail:* `amri@huji.ac.il`

Massive black holes (BHs) are detected in the centers of many nearby galaxies are linearly correlated with the luminosity of the host bulge (spheroid), the black hole mass being about 0.1% of the stellar mass. In active galaxies, the BH mass is best measured by reverberation mapping (light echo) technique. We and others have shown that in AGNs the BH mass follows the same relation with the luminosity of the host galaxy as in ordinary (inactive) galaxies, with the exception of narrow line AGNs which apparently have significantly lower values of the BH/host mass/luminosity ratio. The BH/bulge ratio is also found to be strongly correlated with the velocity dispersion of the broad line-emitting gas in the active nucleus. However, in the $M_{BH}$-$\sigma^*$ relation the difference between broad- and narrow-line AGNs (in particular NLS1s) seems to be smaller. We review the subject adding recent updates and suggestions.

---

[1] Speaker

## 1. Introduction

Massive BHs have been detected in the centers of most galaxies [12]. Magorrian *et al.* [20] suggested that the BH mass is proportional to the luminosity of the host galaxy or host bulge (equivalently, to their mass) hereafter referred to as the BH/bulge relation, with the BH mass being about 0.006 of the mass of the spheroidal bulge (though with a significant scatter). Laor [19] suggested that the masses of quasar BHs and host bulges (both estimated empirically) follow the Magorrian BH/bulge relation. Wandel [14] has investigated the BH/bulge relation in AGN using reverberation data and virial BH mass measurements for 20 Seyfert 1 nuclei [2], and bulge estimates from Whittle [21] with the de Simien-de Vaucouleurs empirical formula [24], finding that in Seyfert nuclei the BH/bulge ratio has a large dispersion, but on average is significantly lower (by an order of magnitude) than the ratio found by Magorrian et al. for quiescent galaxies, as well as the value estimated by Laor for quasars. Over the following years significantly better data have been obtained:

a. HST data have demonstrated that the BH masses in quiescent galaxies have been overestimated by a factor of ~3–5, being 0.001-0.002 of the stellar spheroid mass [13, 22, 42]. The average correction is shown by the arrow marked "G" in fig. 1.

b. BH-masses of 17 PG quasars have been measured by reverberation mapping [23], giving lower BH masses than those estimated by Laor ("Q" in fig. 1).

c. Bulge magnitudes for the Seyfert galaxies in the Seyfert sample have been estimated [23] using HST imaging and bulge-disk 2D decomposition ("S" in fig. 1). Using these improvements Wandel has demonstrated [17] (fig.1) that most AGN do follow the same BH-bulge relation of quiescent nearby galaxies, with a reduced BH/bulge mass ratio of 0.001-0.002. This confirmed the 1999 result, with the modification that only a subgroup of the Seyferts have lower BH/bulge ratios than quiescent galaxies and quasars, NLS1s and narrow line quasars, which have a significantly lower BH/bulge mass ratio (by a factor of ~10) than broad line AGNs and quiescent galaxies.

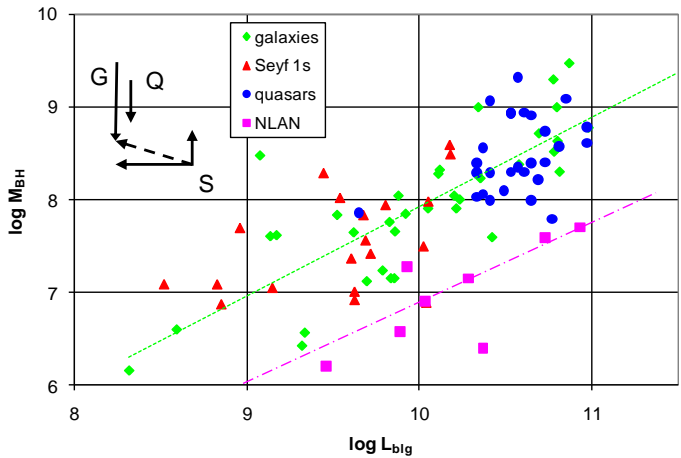

Fig. 1. BH mass vs. host bulge luminosity in quiescent galaxies (green diamonds), Seyfert galaxies (red triangles), quasars (blue circles) and narrow-line AGN (purple squares) and the respective best fits. The arrows show the improvements since 1999 (see text).

A similar conclusion about the lower BH/bulge ratio of NLS1 has been found by Mathur et al. [25] who estimated BH mass of more NLS1s by fitting their X-ray spectra to an accretion disk model calibrated to reverberation mapped BH masses [2], and more recently [41] with HST/ACS bulge estimates and single epoch Hβ BH mass estimates.

In addition to the relationship between BH mass and bulge luminosity, $M_{BH}$ also correlates with the stellar velocity dispersion $\sigma^*$ in the central bulge [13,18,29,30], approximately $M_{BH} \sim \sigma^{*4-5}$. Because of the difficulty to measure the stellar velocity dispersion in NLS it is still unclear whether narrow line AGNs (NLS1 and narrow-line quasars) outlay the "M-$\sigma^*$" relation in quiescent galaxies and broad line AGN [25, 26, 31].

A clear answer to this question could shed light on the cause of the outlaying of NLS1 in the $M_{BH}$-$L_{blg}$ relationship, since if the lower BH/bulge ratio of NLS1s compared with broad line AGNs is, as is often assumed, a result of smaller BHs (or appearing smaller because the BLR has a flattened geometry and is viewed nearly edge on), they should outlay also in the $M$-$\sigma^*$ relation. However, if the reason for their lower $M_{BH}/L_{blg}$ ratios is due to a brighter (intrinsic or apparent) bulge relative to broad line AGN and quiescent galaxies they could appear non-outlaying in the $M$-$\sigma$ relation.

## 2. Reverberation Mapping

AGNs provide a powerful method to estimate the mass of the central BH, based on the response of the broad emission lines to variations in the continuum radiation. Combined with the estimate of the central bulge of the host galaxy (which is more dificult in AGN than in quiescent galaxies), we use this BH mass estimate to investigate the BH-galaxy relation in AGN.

### 2.1 *Schematic structure of AGN*

Analysis of the radiation observed from Seyfert galaxies and quasars indicates several regions on different typical sizes, scaled by the BH mass or the Schwarzshild radius, $R_S = 3 \times 10^{12}$ cm $(M/10^7 M_o)$:

a. the Corona (emitting hard X-rays, at ~$10\, R_S$),
b. the accretion disk and corona (emitting optical, UV and soft X-rays, at 10-100$R_S$)
c. The Broad emission-Line Region (BLR) emitting broad (Doppler width of 1000-10,000 km/s) permitted lines such as Hβ and CIV on scales of $10^3$-$10^5\, R_S$.
d. Farther out, on the sub-kiloparsec scale, there are more extended structures such as the obscuring torus and the Narrow emission-Line Region (NRL), which could play an important role in investigating the inner bulge properties, in particular as a surrogte to the stellar velocity dispersion.

### 2.2 *Emission-Line Response*

In essence, in this method the size of the BLR is estimated from the delay of the variations in the broad emission lines timed by the corresponding variations in optical-UV the continuum which presumably drives the emission lines. A more precise technique known as reverberation mapping or light echo [1] lets us estimate the effective size of the BLR more accurately, given big and frequent enough variability and sampling. The BLR size is estimated from the delay of variations in the emission line flux after corresponding variations in the continuum luminosity. When combined with the velocity of the gas in the BLR measured by the Doppler broadening of the line, the BH mass can be estimated from the Keplerian relation

$$M_{BH} = f v^2 R_{BLR} G^{-1} \qquad (1)$$

where *f* is a factor of order unity, encomapssing the geometry and kinematics of the line emitting gas. It's value may be calibrated[27,28] by studying a large enough sample of AGN with good revereberation data.

### 2.3 *Empirical radius-luminosit relations*

AGNs with light-echo data sufficient for reverberation mapping from multi-year international campaigns include some ~40 Seyfert galaxies & quasars [2,3], a number which has not significantly grown since the early 2000's. These AGN may be used to calibrate an *empirical* expression, the *delay-luminosity* relation. For the reverberation mapped AGN the delay of the Hβ line scales with the non-thermal AGN luminosity as [4,27] $L^{0.5}$. Similar scaling has been demonstrated for the CIV line [5]. This "radius-luminosity relation" makes it possible to estimate the BLR radius (and hence the mass of the central BH) without the extensive observation series required for reverberation mapping. A physical motivation of the *delay-luminosity* relation can be derived using the ionization parameter $\xi = L / 4\pi r^2 n \varepsilon_v c$, relating the delay to the distance of the line-emitting ionized gas from the central BH ($\delta t \sim r/c$) and to the physical conditions in the gas [6-8]. The line emissivity depends strongly on the physical conditions of the emitting gas [9] and in certain emission lines like CIV the emissivity peaks in a diagonal strip in the parameter space, of the form $\xi \sim n^{-1}$ hence $R_{BLR} \sim L^{0.5}$. Combining the continuum and emission line variability data of 17 reverberation mapped Seyferts, Wandel, Peterson and Malkan [2] calibrated a coherent (in the sense it uses the Doppler widening of the same line used to estimate the radius) "M-M relation" between the BH mass measured by reverberation mapping, $M_{rev}$, and the mass estimated by the photo-ionization parameter, $M_{ph}$ (fig. 2).

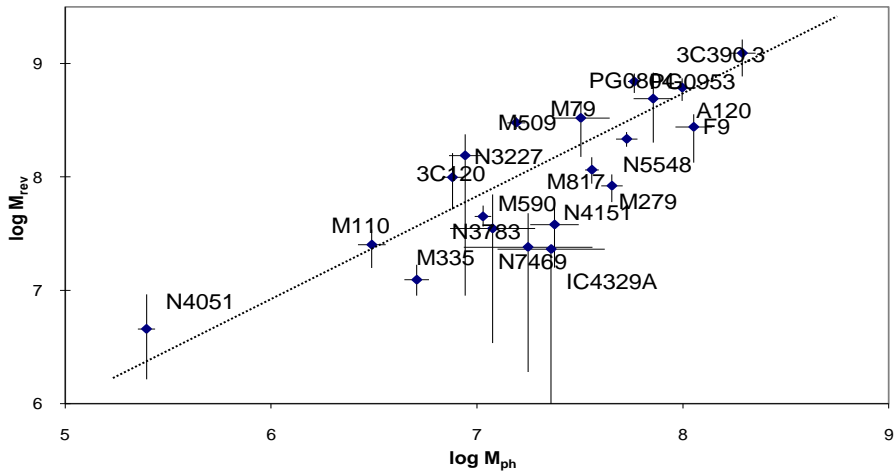

Fig. 2. Correlation between the BH mass measured by reverberation mapping, $M_{rev}$, and the mass $M_{ph}$ estimated by the photo-ionization parameter combined with the Doppler widening of the Hβ line in a sample of Seyfert galaxies with reverberation data.

### 2.4 *Kepler's Signature in Black Holes of AGN*

A multi-year extensive international monitoring campaign produced high quality data for reverberation mapping of NGC5548. Peterson and Wandel [10,11] demonstrated the stratified structure of the BLR by showing that different lines have different time lags ($\tau$) and different Doppler widths (*v*), (fig. 1). This is consistent with a virial radius-velocity relation $v \sim r^{-1/2}$ over a range of 2-30 light days from the center (250-4000 Schwarzshild radii for the estimated BH mass of NGC5548, $6 \times 10^7 M_o$), making a strong case for a central BH.

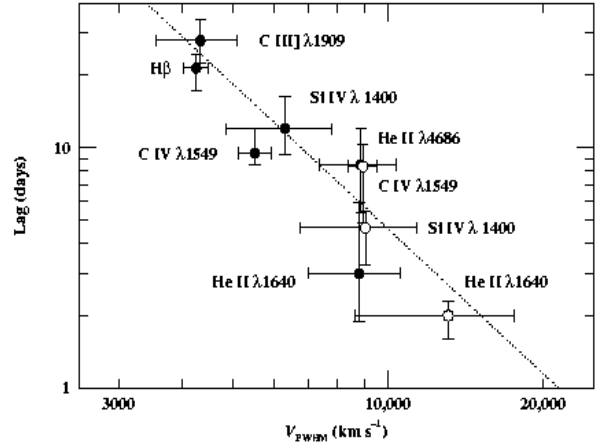

Fig. 3. Time lag vs. Doppler width for broad emission lines in NGC 5548, consistent with the virial relation $v \sim r^{-1/2}$.

## 3. The Black Hole - bulge relation in Active Galactic Nuclei

Using the combined sample of Seyfert 1 galaxies and quasars with reverberation data it has been confirmed [15,17] that AGNs follow the same $M_{BH}$-$L_{bulge}$ realtion found in quiescent galaxies. However, Narrow Line Seyfert 1 galaxies, as well as narrow line quasars appear to have a significantly lower ratio of $M_{BH}/L_{bulge}$ [16,17] (fig. 4)

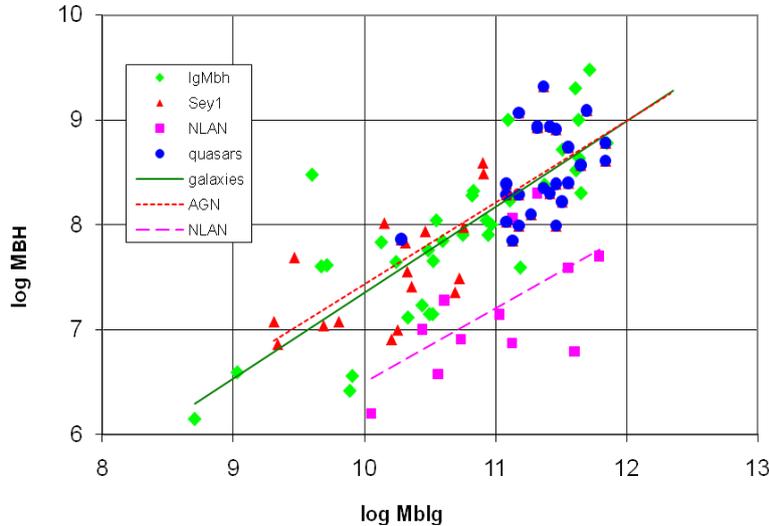

Fig. 4. BH mass vs. host bulge mass in quiescent galaxies (green diamonds), Seyfert galaxies (red triangles), quasars (blue circles) and narrow-line AGN (purple squares) and the respective best fits.

Wandel [17] has also demonstrated that the lower $M_{BH}/L_{blg}$ ratio of narrow-line AGNs compared with broad line AGNs does not reflect two separate population but is rather a continuous distribution in line width (fig 5): the BH/bulge ratio is correlated with the broad emission line width (or the velocity of the line emitting gas):

$$M_{BH}/L_{blg} \sim v^2 .$$

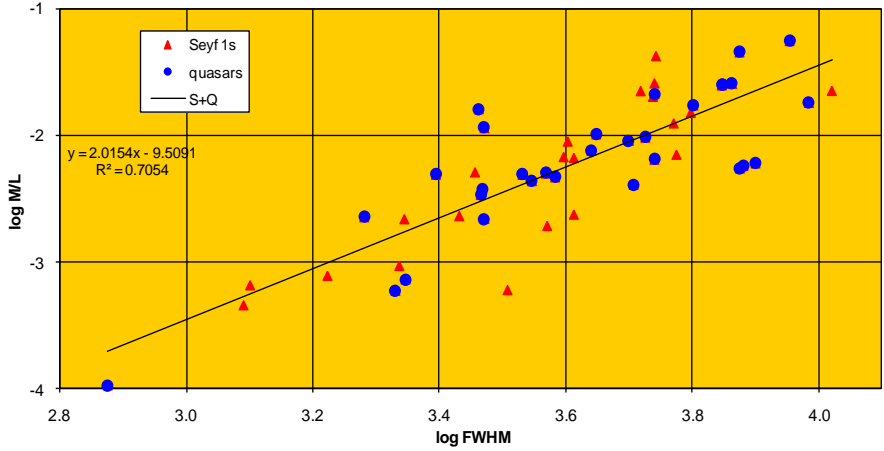

Fig. 5. BH mass to. host bulge luminosity ratio (in solar units) in Seyfert galaxies (red triangles) and quasars (blue circles) and the best fit. Narrow line AGN are not outstanding.

This result is independent of the factor $v^2$ contained in the viral expression (1), as it is the result of the best fit to the data, which could have any power of the velocity. This correlation also implies a new, independent relation between the active nucleus and the host bulge: the size of the Broad Emission-line Region scales with the bulge luminosity (and mass) of the host galaxy,

$$r_{BLR} \sim L_{blg}.$$

Using the empirical radius-luminosity relation we find that the AGN continuum luminosity scales with the bulge luminosity

$$L_{AGN} \sim L_{blg}^2$$

These relations do not show a difference between ordinary and Narrow-line AGNs.

## 4. NLS1s in the $M_{BH}$-$\sigma^*$ relationship

A possible explanation to the location of NLS1s in the $M_{BH}$-$L_{blg}$ plane is that they have smaller BH masses than their broad line counterparts. This effect could be intrinsic or apparent. The latter possibility could be the case if the BLR has a flattened geometry, and is viewed at a low inclination to the line of sight. In that case, the lower $M_{BH}/L_{blg}$ of NLS1 could be an inclination effect (flattened BLR viewed nearly face on). The BH mass of NLS1 measured by assuming isotropic geometry would be under-estimated. In that case, NLS1s should also fall below in the $M_{BH}$-$\sigma^*$ relationship (unless the bulge somehow conspires to show a similarly reduced stellar velocity dispersion as the BLR). It is still unclear whether Narrow Line Seyfert 1 galaxies are consistent with the $M_{BH}$-$\sigma^*$ relation of quiescent galaxies and broad line AGNs. Early works [25,26] show that Seyfert galaxies (including some NLS1s) are consistent with the quiescent galaxy $M_{BH}$-$\sigma^*$ relationship. Botte *et al.* [38] find that NLS1s do lie by an average factor of ~2-3 below the $M_{BH}$-$\sigma^*$ relationship of broad line AGNs and quiescent galaxies. The recent work of Woo *et al.* [31], who have measured a number of Seyferts with very low-mass BHs, may imply that the BH mass of the low $M_{BH}$ and low $\sigma^*$ end has a factor of ~3 lower $M_{BH}/\sigma^*$ than the high end counterpart. This is consistent with Wandel's result [17], which used the Faber-Jackson relation in order to estimate stellar velocity dispersion of additional NLS1 galaxies and narrow line quasars, finding that most narrow-line AGNs lay 0.5 dex below the quiescent galaxy and broad line AGN $M_{BH}$-$\sigma^*$ relationship (fig. 6).

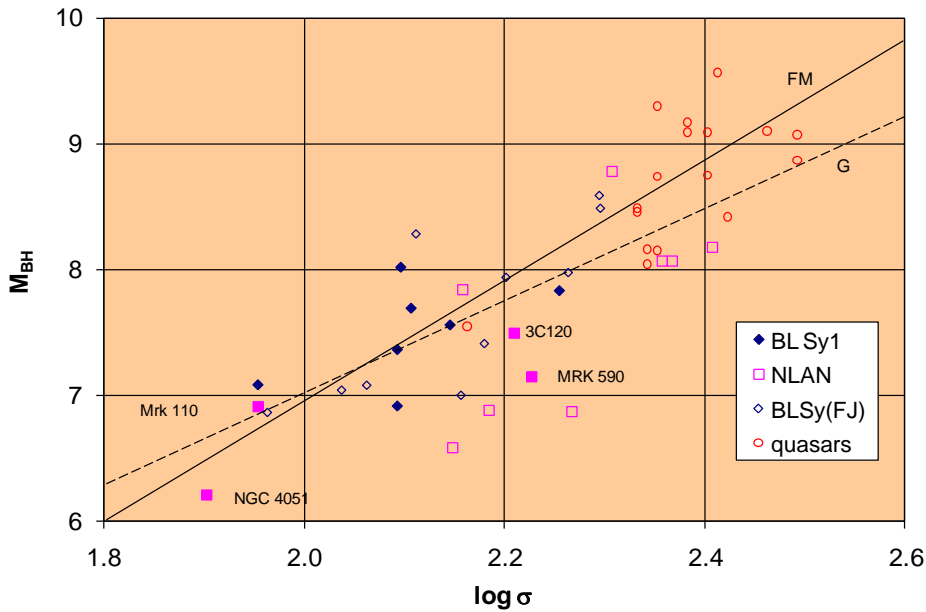

Fig. 6. Black hole mass of AGNs plotted against the stellar velocity dispersion. Blue diamonds are broad line Seyferts, pink squares denote NLS1s and red circles denote quasars. Solid symbols are Seyferts with measured $\sigma^*$, open symbols denote Seyferts with $\sigma^*$ estimated from the Faber-Jackson relation. The diagonal lines are best fits of quiescent galaxies

Because of the difficulty of a precise measurement the stellar velocity dispersion in NLS1 host bulges (due to the strong FeII emission outshines the absorption lines typically used to determine stellar velocity dispersion), many authors have used the O[III] line as a surrogate, shown to obey the $M_{BH}$-$\sigma^*$ relationship [32,33]. Using BH-mass estimates deduced from Einstein X-ray variability of quasars and Seyfert 1 galaxies, Wandel and Mushotzky [34] have demonstrated that the mass estimated from the O[III] narrow line kinematics is well correlated with the central BH mass. This result may be considered as a precursor of the $M_{BH}$-$L_{blg}$ and $M_{BH}$-$\sigma^*$ relationships in AGNs, as the mass within the NLR is likely to be related to the mass of the inner part of the host bulge. Different authors find different answers to the question of whether NLS1 galaxies outlay [16,35-37] or are consistent with [38-40] the $M_{BH}$-$\sigma$(OIII) relationship for broad line AGN, depending on the samples taken and the treatment of the O[III] profile and the NLR kinematics (correction for outflow).

**Summary**

NLS1s seem to have lower BH masses than broad line AGNs for a given host bulge luminosity and mass, and possibly also for a given stellar velocity dispersion $\sigma^*$ or narrow line width $\sigma$(OIII). If present in the $M_{BH}$-$\sigma^*$ relationship, the effect is smaller (0-0.5 dex) than in the $M_{BH}$-$L_{blg}$ relationship (1 dex). This difference between the location of NLS1s with respect to the two relationships could arise because the mass and luminosity of the host bulge is an integral, extended property, while the stellar velocity dispersion is more dominated by the inner part of the bulge. As suggested previously [14,16], the lower BH/bulge ratio of narrow-line AGNs (in particular NLS1s) could indicate that they are in an earlier stage of the AGN phenomenon.